\documentclass{iaus}
\usepackage{graphicx}
\title[Small-scale structure and dynamics of the lower solar atmosphere] %% give here short title %%
{Small-scale structure and dynamics of the lower solar atmosphere} 

\author[S. Wedemeyer-B\"ohm \& F. W\"oger]   %% give here short author list %%
{Sven Wedemeyer-B\"ohm$^1$
\break \and Friedrich W\"oger$^2$}
\affiliation{$^1$Institute of Theoretical Astrophysics, University of Oslo, P.O. Box 1029 Blindern, N-0315 Oslo, Norway; 
email: sven.wedemeyer@astro.uio.no\\[\affilskip]
$^2$National Solar Observatory at Sacramento Peak, P.O. Box 62, Sunspot, NM 88349, USA;
\break email: fwoeger@nso.edu}

\pubyear{2008}
\volume{247}  %% insert here IAU Symposium No.
\pagerange{1--8}
\date{October 24th, 2007}
\jname{ WAVES \& OSCILLATIONS IN THE SOLAR ATMOSPHERE:
HEATING AND MAGNETO-SEISMOLOGY}
\editors{Erd\'elyi, R., ed.}
\begin{document}

\maketitle

\begin{abstract}
The chromosphere of the quiet Sun is a highly intermittent and dynamic phenomenon. 
Three-dimensional radiation (magneto-)hydrodynamic simulations 
exhibit a mesh-like pattern of hot shock fronts and cool expanding post-shock regions
in the sub-canopy part of the inter-network.  This domain might be called 
``fluctosphere''. 
The pattern is produced by propagating shock waves, which are excited at the top of the 
convection zone and in the photospheric overshoot layer. 
New high-resolution observations 
reveal a ubiquitous small-scale pattern of bright structures and dark regions 
in-between. Although it qualitatively resembles the picture seen in models,  
more observations -- e.g. with the future ALMA --  are needed for thorough 
comparisons with present and future models. Quantitative comparisons demand for 
synthetic intensity maps and spectra for the three-dimensional (magneto-)hydrodynamic 
simulations. The necessary radiative transfer calculations, which have to take into account 
deviations from local thermodynamic equilibrium, are computationally very involved so 
that no reliable results have been produced so far. 
Until this task becomes feasible, we have to rely on careful qualitative comparisons of simulations and 
observations. Here we discuss what effects have to be considered for such a comparison.
Nevertheless we are now on the verge of assembling a comprehensive picture of the solar chromosphere 
in inter-network regions as dynamic interplay of shock waves and structuring and guiding 
magnetic fields.  
\end{abstract}

\section{Introduction}

The chromosphere
 of the quiet Sun -- a story full of misunderstandings. 
Apart from the ongoing controversy concerning the heating mechanism 
\cite[(e.g., Fossum \& Carlsson 2005)]{2005Natur.435..919F)}, many details
 of the small-scale structure of the chromosphere
of inter-network regions are still unknown. Already the term 
``chromosphere''\footnote{\cite[Rutten (2007, and references therein)]{2007ASPC..368...27R} 
uses the term ``clapotisphere'' for the 
shock-dominated subcanopy domain in inter-network regions, whereas his 
``chromosphere'' refers to the fibrilar structure  visible in H$\alpha$ only. 
As ``clapotisphere'' stands for standing waves, we here introduce the term 
``fluctosphere'' instead (fluctus = latin for ``wave'').}
is a frequent source of misunderstandings .
Certainly the large variety of phenomena observed 
\cite[(see, e.g., Judge 2006; Rutten 2006, 2007)]{2006ASPC..354..259J,2006ASPC..354..276R,2007ASPC..368...27R}
created a complex puzzle and sometimes apparent contradictions.
For instance, the observed UV emission implies high temperatures, whereas 
the existence of carbon monoxide lines point at much cooler gas 
\cite[(Ayres 2002)]{ayres02}. 
New high-resolution observations -- as reported here -- 
show a highly dynamic and intermittent pattern that cannot be 
explained with the classical semi-empirical models by 
\cite[Vernazza \etal\   (1981, VAL)]{val81} and \cite[Fontenla \etal\   (1993, FAL)]{fal93}. 
Rather a time-dependent three-dimensional model is mandatory. 
A self-consistent model that can fulfill all observational constraints
would be most valuable for summarising the many faces of the chromosphere, 
indicating and understanding the most relevant 
processes. Chromospheric heating is a central issue as it has 
important implications for the atmospheres of other stellar types. 

Here we report on some advances of detailed radiation magnetohydrodynamic 
simulations in comparison with new high-resolution observations. 
Some crucial aspects of such comparisons -- which often result in misunderstandings --
are discussed.
%---------------------------------------------------------------------------------
\section{Observations of the chromosphere}
\label{sec:ibis}

\begin{figure}
  \centering
  \includegraphics[width=10.5cm]{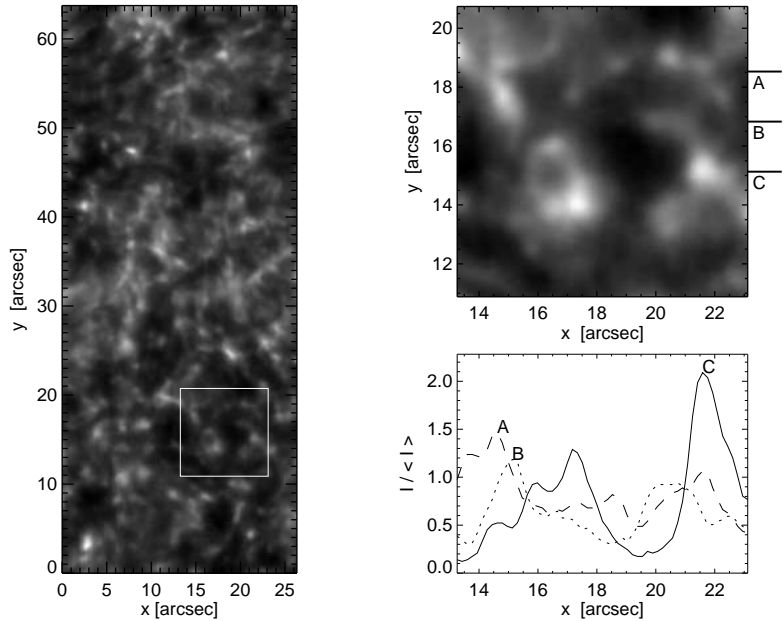}
  \caption{Single IBIS filtergram for the line core of 
  the Ca\,II line at $\lambda = 854.2$\,nm (left) and the 
  close-up of the inter-network region (upper right) marked by
  a white square in the left image. The ticks at the right border
  mark the y positions of the three intensity profiles along the 
  x-axis, which are shown in the lower right panel. 
    }\label{fig:obs}
\end{figure}

The InterferometricBIdimensional Spectrometer 
\cite[(IBIS, Cauzzi \etal\   2007, and references therein)]{2007arXiv0709.2417C}
at the Dunn Solar Telescope of the National Solar Observatory at Sacramento 
Peak is used to scan through the Ca\,II infrared line at $\lambda = 854.2$\,nm
in (2D) spectropolarimetric mode. 
Images at 17 wavelength positions in the line wing and in the 
line core are taken, resulting in a overall cadence of 28\,s. 
The field of view is 26.5"\,$\times$\,64"; the pixelscale 
is 0.17"/px. 
Channels for continuum, G-band, and H$\alpha$  are used simultaneously 
in addition to IBIS. 
 
The Ca\,II line core image for at $\lambda = 854.2$\,nm in 
Fig.~\ref{fig:obs} features a bright mesh-like pattern with dark 
regions inbetween. The spatial scales of the pattern are similar to 
the granulation. Next to the known magnetic network cells 
also a small-scale pattern is visible in the inter-network regions 
(see upper right panel). It exhibits ring-like structures and 
very short-lived bright points. The profiles (lower right panel) show 
a large intensity variation and very small minimum values. 
Image sequences reveal that the small-scale pattern is evolving 
much faster than the granulation and reversed granulation in the 
photosphere below. In particular the bright points only ``flash up''
for a short moment. These new observations support the results reported by 
\cite[W\"oger \etal\   (2006)]{2006A&A...459L...9W}. 
See also \cite[Cauzzi \etal\   (2007)]{2007arXiv0709.2417C}, 
\cite[Tritschler \etal\   (2007)]{2007A&A...462..303T}, and 
\cite[Reardon \etal\   (2007)]{2007ASPC..368..151R}. 
We again interpret the observation as 
manifestation of the interaction of propagating shock waves. The bright 
points are in this picture the collision points of neighbouring wave fronts.

%---------------------------------------------------------------------------------
\section{Numerical simulations}
\label{sec:sim}

\begin{figure}
  \centering	
  \includegraphics[width=0.8\textwidth]{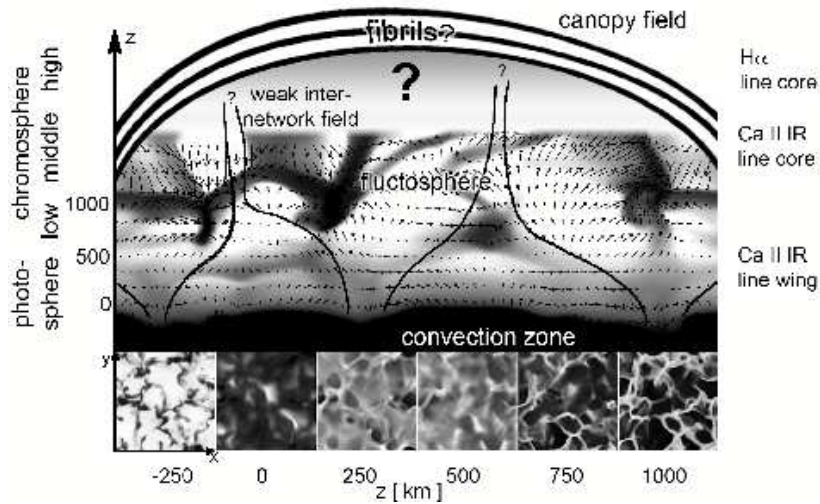}
  \caption{\textit{Upper panel:} Schematic structure of the lower atmosphere 
  in quiet inter-network regions. Velocity field (arrows) and gas temperature 
  (color-coded: white=cool, black=hot) based on the model by 
  \cite[Wedemeyer \etal\   (2004, W04)]{wedemeyer04a}. The lines represent magnetic 
  field lines, forming a canopy (thick) and a weak-field ``small-scale canopy'' 
  (thin) in the inter-network region below.
  On the right the rough (anticipated) formation 
  height ranges of some diagnostics are indicated. 
  \textit{Lower panels:} Horizontal cross-sections at different heights from the model by 
  W04.
  Integral components of the low inter-network atmosphere are the granulation at the bottom 
  of the photosphere ($z \sim 0$\,km), the reversed granulation produced by convective 
  overshooting ($z \sim 250$\,km), a layer with little fluctuations near the height of the 
  classical temperature minimum ($z = 500$\,km), and the fluctosphere ($z > 700$\,km) 
  produced by upward propagating and interacting shock waves.
  The structure of the blank layer marked with ``?'', i.e. the interface between 
  fluctosphere and canopy domain,  is still poorly known as it demands for 
  a sophisticated non-LTE modelling. 
  Please note that the temperature in the lower panels is scaled individually. The 
  variation at $z = 500$\,km is much smaller than in the other layers. 
    }\label{fig:sim}
\end{figure}

The 2D/3D numerical simulations considered here all comprise a small part of 
the surface-near layers and vertically extent from the upper convection zone to 
the middle chromosphere. 
The hydrodynamical model(s) by \cite[Wedemeyer \etal\   (2004)]{wedemeyer04a}, 
computed with CO$^5$BOLD \cite[(Freytag \etal\   2002)]{cobold}, 
feature(s) a small-scale chromospheric pattern, which consists of hot shock fronts 
and intermediate cool post-shock regions (see Fig.~\ref{fig:sim}). It is caused by 
the propagation and interaction of shock waves that are excited in the layers below. 
Due to adiabatic expansion of the post-shock regions the gas temperature reaches 
values down to 2000\,K in the model chromosphere. 
An obvious example of a post-shock region can be seen in the left part of 
Fig.~\ref{fig:sim}. The corresponding strong shock front ``collides'' with the 
neighbouring front to the left, compressing the gas in the region in-between and 
rising its temperature.  
The typical spatial scale is similar to the granular one. The pattern changes 
dynamically on typical time scale of $\sim 30$\,s.
The magnetohydrodynamic model by \cite[Schaffenberger \etal\   (2006)]{2006ASPC..354..345S} 
and also the models by 
\cite[Skartlien \etal\   (2000)]{skartlien00c} and 
\cite[Hansteen \& Gudiksen (2005)]{2005ESASP.592E..87H}
are very similar concerning structure and dynamics. 
Also the magnetic field in the model chromospheres is highly dynamic. 
A look at horizontal cross-sections at different heights
\cite[(see, e.g., Fig.~1 in Schaffenberger \etal\   2006)]{2006ASPC..354..345S} 
implies that the chromospheric field in inter-network regions is much weaker 
($|B| < 50$\,G) than the photospheric one but evolves much faster.
The compression and expansion induced by the ubiquitous propagating 
shocks also play an important role for the small-scale structure of the 
weak inter-network field \cite[(see Fig.~4 in Schaffenberger \etal\   2005)]{2005ESASP.596E..65S}.
Figure~\ref{fig:sim} shows a possible combination of the resulting weak field 
and an overlying stronger magnetic canopy, which is rooted in the network boundaries. 
A similar picture can be seen from other new simulations, e.g., 
by \cite[Leenaarts \etal\   (2007, see Fig.~1 therein)]{2007A&A...473..625L}. 

\label{sec:hion}
The simulations and also the observations presented in the previous sections 
both indicate that the chromospheric inter-network regions of the quiet Sun 
are highly dynamic with inhomogeneities on small temporal and spatial scales. 
This behaviour, which was already suggested by \cite[Carlsson \& Stein (1995)]{carlsson95}, 
certainly has important consequences for the modelling of physical processes.  
An important example is the ionisation of hydrogen. 
Recent 2D and 3D non-equilibrium simulations by 
\cite[Leenaarts \& Wedemeyer-B\"ohm (2006)]{leenaarts06b} 
and \cite[Leenaarts \etal\   (2007)]{2007A&A...473..625L}
confirm the earlier conclusions by \cite[Carlsson \& Stein (2002)]{carlsson02} 
that the ionisation degree and the electron density are fairly constant 
over time and space and tend to be at values set by hot propagating shock waves. 
In contrast, the ionisation degree varies by more than 20 orders of magnitude 
between hot, shocked regions and cool, non-shocked regions when assuming 
instantaneous equilibrium. 
Obviously such deviations from equilibrium must be taken into account 			 
for the physical description of the dynamic solar chromosphere. 
Another important consequence is that the cool intermediate phases leave 
almost no trace in the ionisation degree, which certainly complicates the
observational proof of their existence.

%---------------------------------------------------------------------------------
\section{Comparison of simulations and observations}

On the way towards a comprehensive model of the solar atmosphere, 
detailed comparisons between observations and numerical simulations 
are needed to check if and what physical ingredients are missing and what 
aspects are already modelled realistically. In the case of the chromosphere, 
particular attention has to be paid to the energy 
balance and the gas temperature amplitudes. 
\cite[Kalkofen (2003b, 2005)]{2003SPD....34.1101K, 2005ESASP.560..695K}  
claims that the temperature amplitudes in the dynamical 
models are too large compared to ``modeling based on the emergent spectrum'', where 
the latter refers to semi-empirical models by VAL and FAL. Their models only show  
``modest fluctuations ($\delta T \sim 300$\,K)''  \cite[(Kalkofen 2003b)]{2003SPD....34.1101K}.  
\cite[Kalkofen (2003a)]{2003ASPC..286..443K} argues that the assumption of static atmospheres by 
VAL and FAL is ``justified by the small temperature differences between 
the models which remain below about 5\,\% of the average temperature.'' 
This argument is obviously misleading as it is only 
the difference between averages 
(of different atmospheric regions/brightness components represented by individual VAL models), 
which consequently refers to variations on very large spatial scales and should {\em not} 
be interpreted as the temperature fluctuation introduced by the propagation of waves 
on small spatial scales (say $\Delta x < 2000$\,km).
The existence of strong temperature gradients on small spatial 
scales does {\em not} contradict the models of VAL and FAL
if these are interpreted as average stratifications only. Rather the dynamical models 
can produce a VAL-like average chromospheric temperature rise when 
giving a higher weight to the large temperatures in shock waves 
\cite[(Carlsson \& Stein 1995; Wedemeyer \etal\   2004)]{carlsson95, wedemeyer04a}.

\subsection{Gas temperature and emergent intensity -- A few words of caution}

One must not quantitatively compare observed intensities with the gas temperature 
in a model chromosphere. In particular, just estimating gas temperatures by eye 
from horizontal cross-sections at different heights 
(say with a $\Delta z = 250$\,km) is certainly an inadequate approach. 
Gas temperature cannot directly be interpreted as intensity and vice versa because 
(i) the intensity does not originate from a fixed infinitesimal thin layer but from 
an extended height range and 
(ii) (at least in case of the chromosphere) 
``the convenient approximation of LTE no longer holds for the calculation of radiative 
energy transfer'' \cite[(Kalkofen 2004)]{2004IAUS..219..115K}. 

A look in any textbook on radiative transfer clearly shows that opacity and source 
function do not only depend on the gas temperature but also on parameters as, e.g., 
gas and electron density, chemical composition and ionisation stage of the gas. 
In particular the electron density and ionisation degree are subject 
to deviations from equilibrium in most of the atmosphere (see Sect.~\ref{sec:hion}) 
and thus cannot be 
described in LTE. The chromospheric gas density is modulated by 
compression and expansion due to propagating waves, while, e.g., the population densities 
of atomic 
energy levels depend directly on the (non-local!) radiation field itself. Even in 
a simple case the emergent intensity is always the result of the 
integration of contributions along the line of sight. 
It consequently represents the integrated (thermodynamic) conditions of an extended formation 
height range and {\em not} the local conditions at a particular height. In an 
inhomogeneous intermittent chromosphere the true thermal structure is thus 
obscured by an intrinsic ``smearing'' along the optical depth~$\tau$. 

A comparison of observations and models is thus complicated by the fact that 
the gas temperature can be investigated at individual geometrical heights in the 
models, whereas observed intensity always originates from an extended formation height 
range. In some cases the corresponding contribution functions even have multiple peaks 
at different heights, making the use of the term ``formation height'' problematic.
This is an intrinsic physical problem that cannot be removed. One can only try to 
minimize the effect by choosing a suitable diagnostic 
(with small and well-defined variation in formation height) in combination with the 
detailed comparison by means of non-LTE radiative transfer calculations based on 
forward modelled simulations.

In addition the observations are technically limited by the attainable resolution 
in angle $\Delta \alpha$, time $\Delta t$, and wavelength $\Delta \lambda$:
\begin{enumerate}
\item Atmospheric seeing and instrumental effects limit the spatial scales 
accessible in observations and smear out features on small scales.
\item The formation height ranges vary significantly in 
time and space so that intensities refer rather to corrugated surfaces of 
optical depth instead of plane horizontal cuts. 
\item A limited spectral resolution, e.g. when using a broad band filter, 
essentially mixes together intensity contributions from different height ranges, 
effectively smearing out the small-scale structure of the atmosphere. 
For very narrow filters Doppler-shifts have to be considered. 
\end{enumerate}

In contrast to the unavoidable intrinsic smearing in $\tau$, advances in observational 
technique have the potential to reduce the smearing in $\alpha$, $t$, and 
$\lambda$. The combination of these effects, however, can prevent the detection 
of a cold and diluted post-shock region behind a shock wave with current and past instruments, 
as such regions are quite small and exist for a short time only. 
Consequently empirically derived temperature amplitudes bear the danger 
of being systematically too low. An example might be the low values suggested by 
\cite[Avrett \etal\   (2006)]{2006A&A...452..651A} and 
\cite[Avrett (2007)]{2007ASPC..368...81A}, which rely on SUMER observations
\cite[(Wilhelm \etal\   2005)]{2005A&A...439..701W} with limited spatial resolution. 

\subsection{Ca\,II IR line}
For a detailed comparison with the Ca\,II  data presented here 
(see Sect.~\ref{sec:ibis}) synthetic intensity images still need to 
be calculated. This task, although very involved, is currently in progress.
On the observational side, Doppler-shifts move too fast propagating features 
outside of the filter range. This effect can be corrected for as the IBIS scan 
includes the neighbouring wavelength positions at only small delay.
(A detailed publication is currently in preparation.)

Nevertheless the general picture exhibited by observations and simulations 
is astonishingly similar. Both show the 
chromosphere as highly dynamic and intermittent phenomenon. 
A quantitative comparison by means of non-LTE radiative transfer calculations,
however, will always be hampered by the extended formation height range and 
non-equilibrium effects. The conclusions might thus be of limited value for our 
understanding of the quiet Sun chromosphere. 

\subsection{(Sub-)millimetre continua}
A very promising alternative are observations in the (sub-)millimetre
range with the Atacama Large Millimeter Array (ALMA) -- an array of 50 antennae 
with diameters of 12\,m on a plateau at 5000\,m altitude in the Chilean Andes. 
It will commence full operation in 2012. Wavelengths in the 
range of 0.3\,to\,3.6\,mm will be accessible. 
Synthetic brightness temperature maps have been calculated 
by \cite[Wedemeyer-B\"ohm \etal\   (2007)]{2007A&A...471..977W}.
The maps exhibit very similar spatial and 
temporal scales than the original gas temperature in the model chromosphere, 
which is closely mapped with this kind of diagnostic.
When using the non-equilibrium electron densities from the simulation by 
\cite[Leenaarts \& Wedemeyer-B\"ohm (2006, see Sect. 3)]{leenaarts06b} 
as an input for 
intensity synthesis, the formation height range stays on average very similar 
but varies less in the non-equilibrium approach. 
The resulting brightness temperature is even closer to the gas temperature at a 
fixed geometrical height plane, which simplifies the interpretation. 
In either case the formation height increases with wavelength $\lambda$ and 
heliocentric position ($\mu = \cos \theta$) from centre to limb. 
Consequently the sampled layer can more or less be chosen by $\lambda$ and $\mu$, 
facilitating a tomography of the three-dimensional atmospheric structure.

\section{Assembling a new picture of quiet Sun inter-network regions}
Another source of confusion might arise from assigning fixed formation height 
ranges to the different diagnostics (see, e.g., VAL). 
Considering a substantial overlap of formation 
height ranges of the individual diagnostics and a significant variation in space 
and time (caused by ``fluctospheric'' shock waves, see 
\cite[Carlsson \& Stein 1998]{1998IAUS..185..435C}),  
the different observational constraints can be assembled to a 
comprehensive picture of the quiet Sun atmosphere as, e.g., suggested by
\cite[Rutten (2007,]{2007ASPC..368...27R} 
see also Fig.~\ref{fig:sim} in this article).
It might be so that the emission features like the one near 117\,nm 
(S\,I, \cite[Avrett \etal\   2006]{2006A&A...452..651A})  
but also the He\,II line at 164\,nm \cite[(Wahlstrom \& Carlsson 1994)]{1994\textit{ApJ}...433..417W}
are formed above/in the magnetic canopy layer, while the 
strong shock activity, which, e.g., produces bright points observed in the Ca\,II lines, 
is limited to the ``fluctosphere'' regions below the canopy (see Fig.~\ref{fig:sim}). 
The canopy field certainly modifies the properties of upward propagating disturbances 
\cite[(cf., e.g., Carlsson \etal\   1997; Bogdan \etal\   2003)]{1997\textit{ApJ}...486L..63C, 2003\textit{ApJ}...599..626B}, 
causing distinct differences in structure and dynamics of the sub-canopy and the 
canopy domain.  
It is reasonable to assume that the H$\alpha$ line core  tends to be 
formed higher up in the atmosphere than the cores of the Ca\,II lines
\cite[(cf. Langangen \etal\   2007)]{2007arXiv0710.0247L}.
Indeed a closer look at the high-resolution H$\alpha$ line core observations by 
\cite[Rouppe van der Voort \etal\   (2007)]{2007\textit{ApJ}...660L.169R} not only reveals a 
wealth of fibrils covering the inter-network regions but also the wave-generated 
``fluctosphere'' shining through at locations where the fibrils are (partially) 
transparent. 
To some extent the scene reminds of floating spaghetti pushed around by 
boiling water underneath. 
Depending on the structure and strength of the magnetic (canopy) field, 
fibrils can show up in the line cores of the Ca\,IR lines -- 
provided that a very narrow filter is used
\cite[(see, e.g., Fig.~5 by Cauzzi \etal\   2007)]{2007arXiv0709.2417C}.

%---------------------------------------------------------------------------------
\section{Conclusions and Outlook}
State-of-the-art numerical simulations exhibit a highly dynamic  chromosphere, 
which is characterised by propagating and interacting shock waves with 
co-existing hot and cool regions. 
New observations, as presented here, now have a sufficiently high spatial, temporal 
{\em and} spectral resolution to resolve a pattern of bright structures and dark 
regions. As synthetic intensity images for the cores of the prominent Ca\,II 
lines are still missing, these observations can only be compared to the simulations 
on a qualitative basis at the moment. Nevertheless they clearly support the picture 
of the chromosphere as highly dynamic and intermittent phenomenon like 
it was already implied by the pioneering simulations by 
\cite[Carlsson \& Stein (1995, 1998)]{carlsson95, 1998IAUS..185..435C}.
Even \cite[Kalkofen (2004)]{2004IAUS..219..115K} stated that 
``[d]etailed observations show the chromosphere to be highly dynamic''. 
The model atmospheres by VAL and FAL, although certainly very elaborate, 
suffer from the basic assumption of a one-dimensional stratified static atmosphere. 
This assumption is clearly questioned by recent high-resolution observations.  
Strong intensity fluctuations are now observed although one 
still should need be careful with deriving statements concerning the gas temperature.
Frequently it is argued that the amplitudes and minimum values of gas temperature in 
the model atmospheres are not observed and that they are wrong as they do not agree 
with the VAL models \cite[(cf. Kalkofen 2003b)]{2003SPD....34.1101K}.
These arguments obviously do not hold any longer in view of new observational 
results. 
VAL-type atmospheres should thus be considered as qualitative averages, which 
could at best be interpreted as variations on large spatial scales. 
Instead of aiming at an agreement with VAL models, modern 3D radiation 
(magneto-)hydrodynamical simulations must be directly compared to observations.
Certainly the quantitative values of the temperature fluctuations in these models 
(apart from the low to middle photosphere)
still suffer from the simplified treatment of radiative transfer and resulting uncertainties
in the energy balance, which are, however, a necessary compromise in order to keep the 
computations tractable. 

The next steps towards realistic chromosphere models requires 
more work on the modelling of non-equilibrium effects, which are important 
for the energy balance and thus the temperature amplitudes in the chromosphere. 
An efficient non-LTE radiative transfer scheme is a major goal.

Observational emphasis should be given to (i)~continued aiming at a combination of 
high spatial \textit{and} temporal \textit{and} spectral 
resolution, as they are crucial for a meaningful comparison and interpretation, 
 and (ii) the development of new diagnostics as, e.g., the (sub-)millimetre 
continua. Especially the upcoming ALMA might allow a tomography of the solar 
atmosphere, finally revealing details of its three-dimensional structure.

\begin{acknowledgments}
We would like to thank the organisers of the International Astronomical Union 
Symposium 247 for an interesting meeting. 
SWB is grateful to Vilfredo for stimulating input. 
He was supported by the Research Council of Norway, grant 
170935/V30.
We thank M.~Carlsson, R.~Rutten, L.~Rouppe van der Voort, \O.~Langangen,  
 A.~Tritschler, and K.~Reardon for discussions. NSO is operated by the Association of 
Universities for Research in Astronomy, Inc (AURA), for the National Science Foundation. 
\end{acknowledgments}

%\begin{discussion}
%\discuss
%\end{discussion}

\end{document}